\documentclass[twocolumn,showpacs,aps,prl,superscriptaddress,floatfix]{revtex4}
\usepackage{graphicx}% Include figure files
\usepackage{dcolumn}% Align table columns on decimal point
\usepackage{bm}% bold math

\begin{document}

%\preprint{ APS/123-QED }

\newcommand{\BISCO}{Bi$_2$Sr$_2$CaCu$_2$O$_{8+\delta}$}

\title{Magnetic flux jumps in textured \BISCO}

\author{A. Nabialek}
\affiliation{Brockhouse Institute for Materials Research, McMaster University, Hamilton, Ontario L8S 4M1, Canada.}
\affiliation{Institute of Physics, Polish Academy of Sciences, Al. Lotnikow 32/46, 02-668 Warsaw, Poland.}
\author{M. Niewczas}
\affiliation{Brockhouse Institute for Materials Research, McMaster University, Hamilton, Ontario L8S 4M1, Canada.}
\author{H.Dabkowska}
\affiliation{Brockhouse Institute for Materials Research, McMaster University, Hamilton, Ontario L8S 4M1, Canada.}
\author{A.Dabkowski}
\affiliation{Brockhouse Institute for Materials Research, McMaster University, Hamilton, Ontario L8S 4M1, Canada.}
\author{J.P. Castellan}
\affiliation{Brockhouse Institute for Materials Research, McMaster University, Hamilton, Ontario L8S 4M1, Canada.}
\author{B.D. Gaulin}
\affiliation{Brockhouse Institute for Materials Research, McMaster University, Hamilton, Ontario L8S 4M1, Canada.}
\affiliation{Canadian Institute for Advanced Research, 180 Dundas St. W.,Toronto, Ontario, M5G 1Z8, Canada}
\date{\today} %January 22, 2002}

\begin{abstract}
Magnetic flux jumps in textured \BISCO have been studied by means
of magnetization measurements in the temperature range between
1.95 K and T$_c$, in an external magnetic field up to 9 T. Flux
jumps were found in the temperature range 1.95 K - 6 K, with the
external magnetic field parallel to the c axis of the investigated
sample. The effect of sample history on magnetic flux jumping was
studied and it was found to be well accounted for by the available
theoretical models. The magnetic field sweep rate strongly
influences the flux jumping and this effect was interpreted in
terms of the influence of both flux creep and the thermal
environment of the sample. Strong flux creep was found in the
temperature and magnetic field range where flux jumps occur
suggesting a relationship between the two. The heat exchange
conditions between the sample and the experimental environment
also influence the flux jumping behavior. Both these effects
stabilize the sample against flux instabilities, and this
stabilizing effect increases with decreasing magnetic field sweep
rate. Demagnetizing effects are also shown to have a significant
influence on flux jumping.
\end{abstract}


\maketitle

\section{INTRODUCTION}
Magnetic flux jumps are one of the peculiar phenomena of interest in both conventional,
hard type II superconductors and in high temperature superconductors (HTSs).
Studies of flux instabilities in superconducting materials are of interest both from
a basic point of view and also in light of their potential applications. In addition,
the investigation of flux jumps in HTSs are relevant to the understanding the complexity
of the vortex matter in the mixed phase of these materials. It is known that under
appropriate conditions, the critical state of a superconductor may become unstable, leading
to an avalanche-like process, initiated by a small fluctuation of the temperature or
external magnetic field. This process is associated with the sudden puncture of
magnetic flux into the volume of the superconductor with a corresponding increase in
the material's temperature. During this process, the screening current is appreciably
reduced, perhaps  even to zero. From the point of view of the applications, magnetic
flux jumps are problematic as they may drive superconductor into a normal or resistive
state. Flux jumps cause also abrupt changes to the sample dimensions, which may be
observed via magnetostriction measurements \cite{nab97}. Flux jumps phenomenon have been
studied primarily by magnetization measurements, screening experiments and torque
magnetometry (see reference \cite{wip92} for review).

The basic theory appropriate to magnetic flux jumping was developed in late 1960s by
Swartz, Bean and Wipf \cite{wip65,wip67,swa68}. Theoretical analysis usually assumes
the fulfillment of the local adiabatic conditions for the sample, which in turn depends
upon the relation between the thermal (D$_t$) and the magnetic (D$_m$) diffusivity of
the material. If {D$_t$$\ll$D$_m$}, the local adiabatic conditions for the occurrence
of the flux jump process are assumed to be satisfied. Flux jumps are associated with the
diffusion of magnetic flux into the superconductor. The diffusion time $\tau$$_m$ of
magnetic flux, is inversely proportional to magnetic diffusivity, i.e.
{$\tau$$_m$$\sim$1/D$_m$}. Similarly, the thermal diffusion time $\tau$$_t$ is inversely
proportional to thermal diffusivity, {$\tau$$_t$$\sim$1/D$_t$}.
If the thermal diffusion time is significantly longer than the characteristic time for
the process, the conditions of this process are considered to be locally adiabatic.
In the case of  magnetic flux jumps, these conditions are satisfied when
{$\tau$$_t$$\gg$$\tau$$_m$} or {D$_t$$\ll$D$_m$}.

The stability criteria of the critical state of hard type II superconductors may be
obtained by analysis of a loop of several interconnected processes, as was discussed
for example by Wipf \cite{wip92} (see also Fig.2 in reference \onlinecite{wip92}).
A small thermal fluctuation, $\Delta$T$_1$, causes an appropriate decrease in the
critical current density. This in turn decreases the screening current of the
superconductor, allowing some additional magnetic flux to enter the volume of the
sample. The additional flux causes heat dissipation, which generates an additional
increase of the temperature of the superconductor by amount of $\Delta$T$_2$.
If $\Delta$T$_2$$>$$\Delta$T$_1$ an avalanche-like process in form of a flux jump is
induced. The range of temperature and magnetic field for which flux jumps occur is
determined by two parameters. The first one is the instability field B$_{fj}$. In the
adiabatic approximation, and for an infinite slab geometry sample, the instability field
for the first flux jump (after cooling the sample in zero magnetic field) is given by the
formula:

\begin{eqnarray}
 B_{fj1} &=& \sqrt{ \frac {2\mu_{0}cJ_{c}} {-dJ_{c}/dT} }
\label{eq:one}
\end{eqnarray}

where c is specific heat, J$_c$ is critical current density and, $\mu$$_0$ and T are
the magnetic permeability of vacuum and temperature, respectively. This theory assumes
that J$_c$ is independent on the magnetic field, and B$_{fj1}$ are measured after
cooling the sample in zero magnetic field.

At sufficiently low temperatures, both the
specific heat of the superconductor and the instability field for the first flux jump
B$_{fj1}$ increase with temperature. At some higher temperature, the B$_{fj1}$(T)
curve reaches a maximum and then drops to zero at the critical temperature of the
superconductor (T$_c$), because at this temperature the critical current of the
superconductor vanishes. At 4.2 K typical values of B$_{fj1}$ predicted by eq. (1) are
of the  order of 0.1 T. Thus, flux jumping is expected to be an important problem from
the point of view of applications of type II superconductors, as it limits the
performance of these materials in a low temperature regime.

The second parameter affecting the appearance of flux jumps is the critical dimension
of the superconductor, i.e. the minimum sample dimension for which flux jumps occur.
The critical dimension of the sample depends on the sample's shape, its orientation
relative to the external magnetic field and on the relation between
B$_{fj1}$ and the value of the field of full penetration (B$^*$) of the superconductor.
For the case of infinite slab geometry
sample, or for an infinitely long cylindrical sample with an external magnetic field
aligned parallel to its surface (slab sample) or axis (cylindrical sample), the role
of the critical dimension is played by the sample diameter.
The parameter B$^*$ depends on the sample shape, the sample's orientation in the
magnetic field and the field dependence of the critical current density.
To the good approximation B$^{*}$ is proportional to the critical current
density (J$_{c}$) and sample dimensions.

For the case of an infinite slab
or cylinder and under the assumption that J$_{c}$(B)=constant
(the so-called Bean model \cite{bean62}), the field of full penetration can
be calculated using relation B$^*$=$\mu_0$J$_c$d, where d is a half
of the diameter of the infinite slab or the cylinder.
At sufficiently low temperatures
B$_{fj1}$ increases with increasing temperature, whereas B$^*$ decreases. At some
temperature (T$_1$), B$^*$(T$_1$)$=$B$_{fj1}$(T$_1$) and at higher temperatures
B$^{*}$$<$B$_{fj1}$. Assuming that J$_{c}$(B)=constant, the critical diameter of
the slab or cylinder
sample for which $\Phi_{crit}$=2d$_{crit}$, may be determined from the equation
B$^*$(T$_1$)=$\mu_0$J$_c$(T$_{1}$)d$_{crit}$. For a slab (or cylinder) with
diameter smaller than the critical one, no flux jumps occur at any temperature for
any external magnetic field, independent of whether the measurements are made during
the virgin magnetization curve (taken after cooling the sample in zero external
magnetic field) or for other parts of magnetization hysteresis loop. However, when
the critical current density depends on magnetic field, the situation becomes more
complex. In this case, even if there are no jumps in the virgin magnetization curve,
some jumps (so-called "solitary jumps" \cite{mll94}) may appear after reversal of
external magnetic field direction. When the critical current density is a
non-monotonic function of the magnetic field (the so-called "fish tail" or
"butterfly" effect), then  under appropriate conditions so-called "island jumps" may
also be found \cite{cha98}. All these phenomena result from the changes in
magnetic field profile in a superconducting sample caused by the field dependence of
the critical current density as well as by magnetic history. In all cases, however,
some critical dimension of the sample exists and for samples with dimensions
smaller than the critical one, no jumps occur at any temperature or magnetic field.
Thus, by reducing the diameter of the superconductor it is possible to avoid flux
jumping.  This approach is commonly used in producing multifilament superconducting
wires. Flux jumping in conventional superconductors, including the extension of
the basic flux jumping theory to non-adiabatic conditions, has been thoroughly
reviewed in reference \cite{min81}.

Magnetic flux jumps have also been observed in high temperature superconductors
\cite{wip92}. Because of the existence of a critical dimension, flux jumping in
HTSs was observed only in relatively large single crystals or well-textured
polycrystalline samples with high values of the critical current. No such effect
has been observed in ceramics, because the critical dimension in these materials is
limited by the grain size, which is typically very small, of the order of several
microns. Most of the studies of flux jumps in HTSs that have been reported to date
have been carried out on YBa$_2$Cu$_3$O$_{6+\delta}$ \cite{wip92} or
La$_{2-x}$Sr$_x$CuO$_4$ \cite{nab97,mch92,ger93} single crystals, or on highly
textured polycrystalline materials. In contrast, there are only a few reports of
magnetic flux jumping in the Bi$_2$Sr$_2$CaCu$_2$O$_{8+x}$ system.

Early experiments by Guillot and co-workers \cite{gui89}, observed flux jumps in an
assembly of preferentially oriented Bi$_2$Sr$_2$CaCu$_2$O$_{8+x}$ single crystals.
Unfortunately, no systematic studies of this phenomenon were reported.
Magnetic flux jumps were also found in a large polycrystalline
Bi$_2$Sr$_2$CaCu$_2$O$_{8+x}$ flux tube \cite{bur94}. More systematic
observations of flux jumping in BiSCCO system have been reported by Gerber
and co-workers \cite{gerb93}. This work reports flux jumps in a
Bi$_2$Sr$_2$CaCu$_2$O$_8$ sample consisting of a pile of c-oriented
single-crystalline slabs, but only at relatively high magnetic field sweep rates,
above 1 T/s \cite{gerb93}.
Among HTS materials Bi$_2$Sr$_2$CaCu$_2$O$_{8+x}$ is characterized by a strong anisotropy,
much stronger than is the case in La$_{2-x}$Sr$_x$CuO$_4$ or YBa$_2$Cu$_3$O$_{6+\delta}$.
Hence, flux jumps studies of Bi$_2$Sr$_2$CaCu$_2$O$_{8+x}$ may be useful in understanding
the development of the instability process in strongly anisotropic superconductors.
A detailed understanding of the instability process in Bi$_2$Sr$_2$CaCu$_2$O$_{8+x}$ is
also important from the viewpoint of potential applications of Bi-based compounds
in Ag/BiPbSrCaCuO composites.

Many aspects of magnetic flux instabilities in HTSs as well as in conventional
superconductors require further investigation to both elucidate the intrinsic dynamics
of magnetic flux in superconductors, and to enable future applications. The present paper
deals with systematic studies of magnetic flux jumping phenomena in c-oriented, textured,
\BISCO samples at magnetic field sweep rates between 0.06-1.2 T/min.
Specifically, we investigate the temperature dependence, magnetic field sweep rate dependence,
as well as the presence of magnetic flux creep on the occurrence of flux jumping.
The influence of demagnetizing effects on flux jumping is discussed in the framework of
the Brand, Indenbom and Forkl model of the critical state of an infinitely long and
thin superconducting strip in an external magnetic field perpendicular to its surface
\cite{bra93}.

\section{EXPERIMENTAL DETAILS}
A \BISCO polycrystalline boule sample with a circular cross-section of about 6 mm in
diameter was grown by the floating zone technique in a four mirror optical furnace. The
critical temperature of the as-grown material was about 92 K, very close to optimal for
this HTS. From the as-grown boule, a sample of approximate dimensions
4.2x2.2x0.2 mm$^{3}$ was detached by cleaving, such that the shortest edge of the
detached sample was parallel to the c-axis.

X-ray diffraction studies of the cleaved sample show it to consist of a mosaic of several
well-aligned single crystals. The rocking curve showed
that the total misalignment of the c-axes of this mosaic of single crystals was about 5
degree. This sample was used for all further studies of magnetic flux jumps reported
here. Magnetization measurements were carried out using a Quantum Design PPMS-9 system
with the maximum external attainable field of 9 Tesla. The measurements were performed
using the extraction magnetometer option. In this option the sample is moved between
two pick-up coils with constant velocity, whereas the signal from the pick-up coils is
integrated to calculate the magnetic moment of the sample under study. During experiment
the sample was surrounded by a low pressure helium gas (around 0.5 Tr) thus,
similar heat sinking conditions were maintained by crossing from 1.95 K to above 4.2 K.
In all the magnetization measurements reported here, the temperature was varied between
1.95 K and T$_{c}$ and the external magnetic field was changed in sweep mode. The
temperature dependence of magnetization hysteresis loops was measured with constant
sweep rate of 0.3 T/min. In addition, the magnetic field sweep rate dependence of the
flux jumping was studied systematically at a single temperature of 4.2 K. The sweep rates
were adjustable between the maximum and the minimum rates attainable in our system i.e.
between 0.06 T/min and 1.2 T/min respectively.

\section{RESULTS}
Figure 1 shows magnetization hysteresis loops for our sample, with the external
magnetic field parallel  to its c-axis.  This data was obtained for temperatures
between 1.95 K and 6.5 K. All the measurements presented here were performed after
cooling the sample in zero magnetic field. The external magnetic field was then swept
from 0 T to 9 T, back to -9 T and again back to 0 T.  The magnetic flux jumps are
observed in the temperature range from 1.95 K up to 6 K and they are not evident at
6.5 K, as Figure 1 shows. Such magnetization hysteresis loops were also measured above
6.5 K, to temperatures up to T$_c$, but these measurements showed no magnetic flux
jumping in the investigated sample. Similar measurements, carried out with magnetic
field parallel to the sample surface i.e. perpendicular to the c-axis and at the same
sweep rate of 0.3 T/min showed no magnetic flux jumps in the whole temperature range
studied i.e between 1.95 K and T$_c$.

\begin{figure}[t] \centering
\includegraphics[width=0.95\columnwidth]{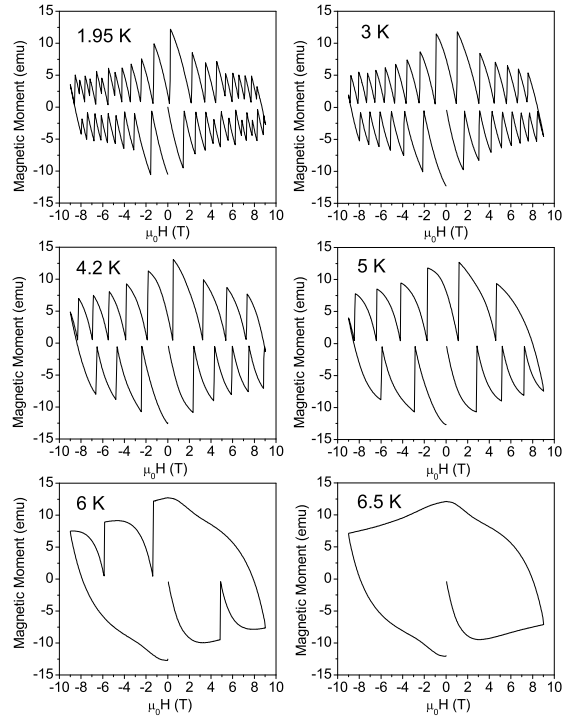}
\caption {Magnetization hysteresis loops taken of the 4.2 x 2.2 x 0.2 mm$^{3}$ \BISCO
textured sample in the temperature range 1.9 K - 6.5 K are shown. The c-axis was parallel to
the external magnetic field and the magnetic field sweep rate was 0.3 T/min.}
\end{figure}

As can be seen in Figure 1, the number of observed jumps decreases with increasing
temperature. Increasing the temperature increases both the value of the field at the
first flux jump, B$_{fj1}$ (see also Fig. 4), as well as an increase in the field
spacing between subsequent jumps. At temperatures above 3 K, all of the observed jumps
are complete, meaning that the magnetization of the sample during a jump drops to zero.
Figure 1 shows also the influence of the magnetic history on flux jumping. This
influence is most clearly seen in the hysteresis loop measured at relatively high
temperatures such as 6 K.   Here one can see that only a single jump occurs within
the first quadrant of the M(H) plot i.e. while the  external magnetic field increases
from 0 T to 9 T. There are no jumps observed in the second quadrant of the M(H) curve,
when the magnetic field is decreased from 9 T to 0 T. Subsequently there are two jumps
present when the external magnetic field is changed from 0 T to -9 T (within the third
quadrant).

The magnetic field sweep rate dependence on flux jumping has been studied at the
temperature of 4.2 K and these results are shown in Figure 2. These measurements were
carried out in the first quadrant of the M(H) hysteresis loop. With increasing sweep
rate, and external magnetic field in the range from 0 T to 9 T, the number of observed
flux jumps increases, whereas the value of B$_{fj1}$ decreases (see also Fig. 6). We
emphasize that, at the lowest attainable sweep rate of 0.06 T/min and the same
temperature of 4.2 K, no flux jumps occurred in our sample. Figure 2 shows all
experimental points recorded. One can see that the observed jumps occur in a time
interval shorter than that between the subsequent experimental points. The minimum
time interval between experimental points is limited in our system to about 5 s.

\begin{figure}[t] \centering
\includegraphics[width=0.95\columnwidth]{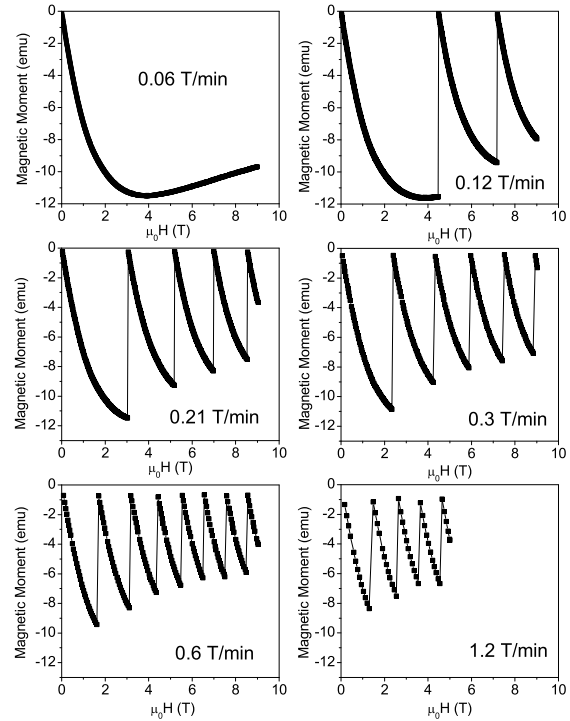}
\caption {The dependence of flux jumping versus the external magnetic field sweep
rate at the temperature 4.2 K is shown. The measurements were taken after cooling the
sample in zero magnetic field with sweep rates 0.12, 0.21, 0.3, 0.6, 1.2 T/min,
respectively. The c-axis of the sample was parallel to the external magnetic field.}
\end{figure}

In order to analyze the influence of flux creep on flux jumps we have measured the
magnetic relaxation of our sample. The following procedure was adopted during the
measurement. The sample was cooled in zero magnetic field to the temperature of 4.2 K,
at which time the external magnetic field was increased from zero to 2 T at a rate of
0.3 T/min. This value of the field, 2 T, was chosen to prevent appearance of flux
jumps, which first occur at this temperature at slightly higher magnetic fields i.e.
at 2.38 T  (see Figures 1,2,4,6). Measurements of magnetic moment were then performed
over a period of one hour and these results are shown in Figure 3. The relaxation of
the magnetic moment is approximately logarithmic in time.  After one hour, about a
10\% decrease from the initial value of magnetic moment is observed in the sample.

\begin{figure}[t] \centering
\includegraphics[width=0.95\columnwidth]{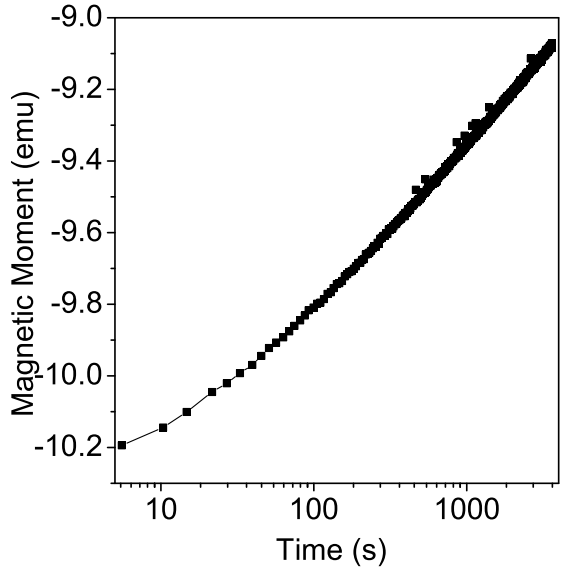}
\caption {The  relaxation of the magnetic moment  at 4.2 K in an external magnetic
field 2 T is shown.  The sample was first cooled in zero magnetic field,  followed by
an increase of the magnetic field to 2 T at a rate of  0.3 T/min. Subsequent to this,
the relaxation of the sample was followed for one hour.}
\end{figure}

\section{ANALYSIS}
The behavior of the magnetization jumps displayed by our \BISCO sample and shown in
Figure 1, is typical for thermally activated flux jumps in type II superconductors
\cite{nab97,wip92,wip65,wip67,swa68,mll94,cha98,min81,mch92,ger93,gui89,bur94,gerb93}.
This data allows one to construct a relation between the
field of the first flux jump, B$_{fj1}$ and temperature, as shown in Figure 4.
The fact that B$_{fj1}$ increases with increasing temperature is consisitent with the
adiabatic theory of flux jumping \cite{wip65,wip67,swa68}, discussed earlier.

\begin{figure}[t] \centering
\includegraphics[width=0.95\columnwidth]{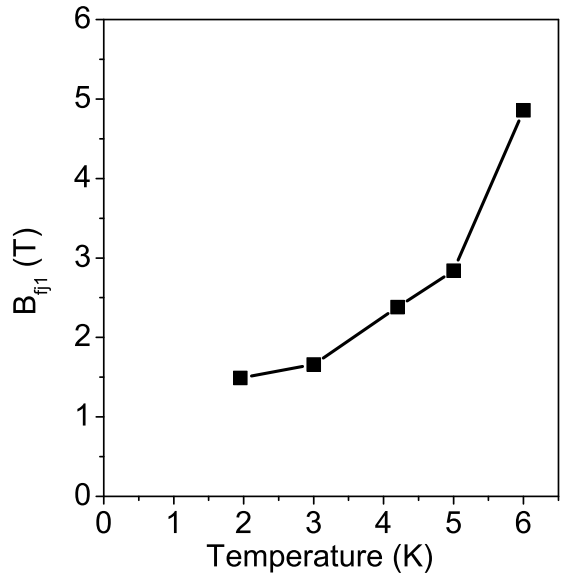}
\caption{The temperature dependence of B$_{fj1}$, determined on the basis of the
data presented in Fig.1 is shown.}
\end{figure}

The present results show that all flux jumps we observe at temperatures higher than
3 K are complete. This indicates that the energy released during the jump is
sufficient to drive the superconductor into the resistive state, which means that this
energy is sufficient to increase the temperature of the sample to the value at which
the critical current density of the superconductor vanishes. It is important to note
that to reduce critical current density to zero in the case of HTSs, it is not necessary
to heat the superconductor above T$_c$ but just above the vortex melting temperature,
where the vortices become virtually unpinned. At lower temperatures
more energy is required to drive superconductor into the resistive state as compared
to higher temperatures.  The temperature increase during a jump depends on the
quantity of energy released during the jump, and on the specific heat of the sample.
At 1.95 K the magnetic field interval between the successive jumps is less than at
higher temperatures. Thus the magnetic energy stored in the superconductor between two
successive jumps at 1.95 K is smaller than it is at higher temperatures.  It may be
that this magnetic energy is insufficient to drive the superconductor into the
resistive state.  Therefore for a number of jumps observed at 1.95 K, the
magnetization of the sample does not drop to zero, as one can see in Figure 1.

The magnetic history dependence on flux jumping, shown in Figure 1, can also be
accounted for by the theory presented in references \cite{mll94,cha98}. In these
references, the temperatures and magnetic field range for which flux jumps occur was
studied in the framework  of the adiabatic theory, assuming different dependences of
the critical current density as a function of the magnetic field. It was found that, if
the critical current density decreases with increasing magnetic field, flux jumps first
appear in the third quadrant of the M(H) plot. Therefore, this quadrant of the M(H)
plot is least stable from the point of view of appearance of flux jumping. Comparison of
the magnetization hysteresis loops at different temperatures shows that as the
temperature is lowered flux jumps appear in the first quadrant and also at the end in
the second quadrant of the M(H) plot. Thus, the second quadrant of the M(H) plot is the
most stable from the point of view of appearance of flux jumping. This behavior of the
flux jumps has been also observed in our sample, as shown in Figure 1
(e.g. temperature 6 K).

\begin{figure}[t] \centering
\includegraphics[width=0.95\columnwidth]{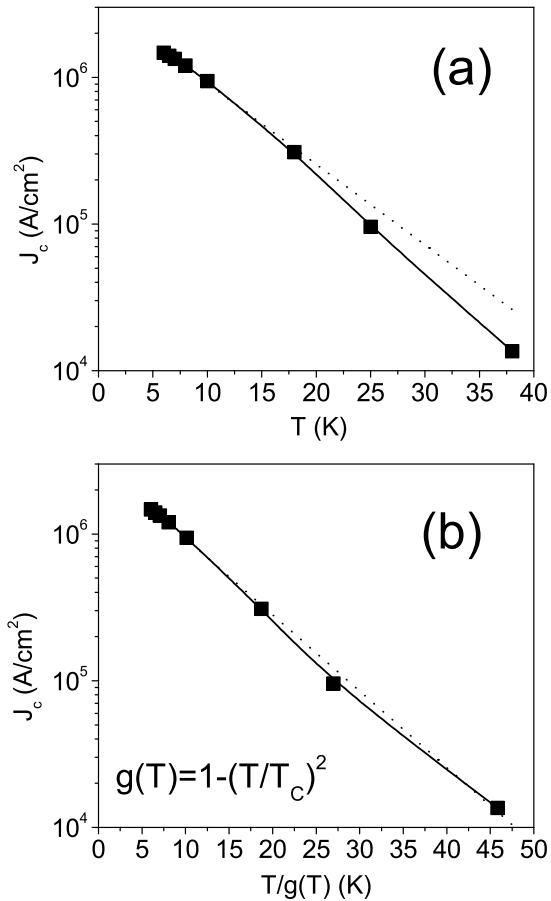}
\caption
{(a) The temperature dependence of the critical current density in the direction
parallel to the ab-plane for zero external magnetic field in logarithmic scale is shown.
The data was obtained on the basis of magnetization hysteresis loops measurements with
an external magnetic field parallel to the c-axis of the  sample.
(b) The same data in a normalized T/g(T) scale, where g(T)=1-(T/T$_{c}$)$^{2}$
and T$_{c}$=92 K. Dotted lines show fit according to the exponential formula
J$_c$(T$^*$)=J$_{c0}$*exp(-T$^*$/T$_0$), see the text for details.}
\end{figure}

In order to quantitatively compare the present experimental results with the predictions
of the adiabatic theory, we have estimated the value of B$_{fj1}$ at 4.2 K.
This requires an estimate of the specific heat as well as temperature dependence of
the critical current density. Determination of the critical current densities on the
basis of magnetization hysteresis loop measurements as well as the stability of the
critical state depends strongly on the distribution of the screening currents in a
superconducting sample. This problem is important in the case of polycrystalline HTSs,
where the grain boundaries may act as the weak links, reducing the critical current
density even by several orders of magnitude. However, the texture of the
polycrystalline samples is of great importance for limiting the critical current.
If the polycrystalline sample consists of a set of well aligned grains, a large current
may be shunt by the substantial common areas between adjacent grains, bypassing the weak
links. This phenomenon is described in terms of the so-called "brick wall" model
\cite{bula92} and it is commonly observed in superconducting thin films as well as
in Ag/BiPbSrCaCuO tapes, which despite their polycrystalline structure are characterized
by high critical current densities. Similar is applicable in the case of our sample,
when an external magnetic field is parallel to the c-axes of the grains. In this case,
screening currents flow within the ab-planes of the grains and they have very large
areas with which to bypass the weak links. Hence, in our further analysis the whole
sample will be treated as a single grain and the effect of the weak links will not be
taken into account. It is also worth noting that after finishing our measurements the
sample was cleaved again into several smaller pieces, each of which possessed almost
the same surface area as the original sample. This suggests that the original sample
had a sandwich-like structure of the weak links, which would not limit the screening
currents in ab-planes, when the external magnetic field is parallel to the c-axes of
the grains.

The temperature dependence of the critical current density in the direction parallel to
the ab-plane is estimated based on the magnetization data and these results are shown
in Figure 5. The widths of the magnetization hysteresis loops were measured at different
temperatures at the zero external magnetic field, and subsequently, a value
of the critical current density was estimated using a formula appropriate to a sample
whose cross-section in the plane perpendicular to an external magnetic field is in
shape of a rectangle, as discussed in reference \cite{gyo89}. As the flux jumps occur
primarily at low temperatures, this procedure cannot be effectively used in this range,
below 6 K. Thus, the temperature dependence of the critical current density was
estimated only for temperatures above 6 K. Figure 5 shows this result between 6 K and
38 K.

As one can see in Figure 5 at zero external magnetic field and at 6 K critical current
densities of the order of 10$^6$ A/cm$^{2}$ were found. These values of the critical
current
density are close to the upper limit of the critical current densities usually observed
in the \BISCO system. This fact is probably connected with the presence of a large number
of structural defects acting as pinning centers. The temperature dependence of the
critical current density at zero magnetic field is roughly exponential. However, it is
difficult to fit a unique exponential formula to the whole range of experimental data
(see Fig. 5a). An exponential temperature (as well as magnetic field) dependence of
the critical current density in superconductors with weak pinning is
expected from flux creep \cite{mchenry94}. A precise analysis of these dependencies
must also reflect the scaling of fundamental pinning related parameters on temperature,
as well as on magnetic field \cite{mchenry91}. To take into account the dependence
of these  parameters on temperature we have rescaled the temperature axis by a function
g(T)=1-(T/T$_c$)$^2$ (see Figure 5b). This function is consistent with  Ginzburg-Landau
theory and is discussed in references \cite{mchenry94,mchenry91}. In the case of HTSs the
g(T) function is usually assumed in the form g(T)=1-(T/T$_{irr}$)$^2$, where T$_{irr}$
is the irreversibility temperature.
At zero external magnetic field we have assumed
T$_{irr}$=T$_{c}$. As one can see in Figure 5, after application of this procedure,
very good fitting of the available experimental data is obtained by using exponential
formula J$_c$(T$^*$)=J$_{c0}$*exp(-T$^*$/T$_0$) where: T$^{*}$=T/g(T),
J$_{c0}$=3*10$^6$ A/cm$^2$ and T$_0$=8.4 K.

\begin{figure}[t] \centering
\includegraphics[width=0.95\columnwidth]{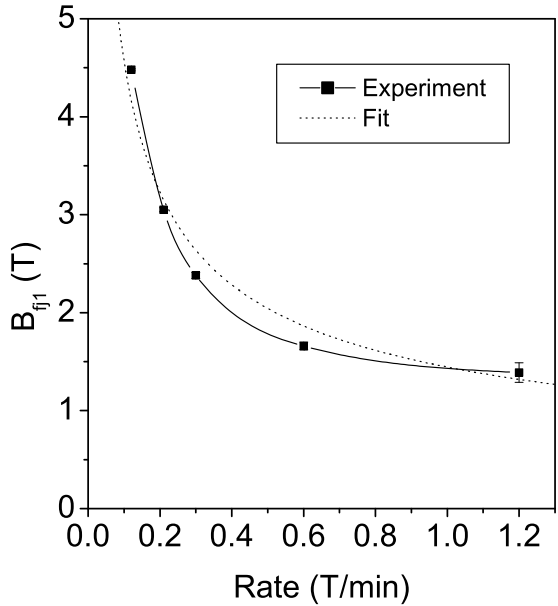}
\caption{The dependence of B$_{fj1}$ on the sweep rate at a temperature of 4.2 K is shown.
 This was determined on the basis of data presented in Figure 2. Experimental data
 are connected by spline (solid curve). The dotted line shows fit according to
 formula (32) from reference \cite{dmints96}.}
\end{figure}

The specific heat of \BISCO at 4.2 K was estimated using
c(T)=$\beta$T$^{3}$ with $\beta$=2 mJ/K$^{4}$mol \cite{cas89,fis88},
from which we get c(4.2 K)=11x10$^2$ J/Km$^3$. Using this value of specific heat,
the above temperature dependence of the critical current density and
formula (1), we estimate B$_{fj1}$(4.2 K)$\approx$0.15 T, a value roughly one order
of magnitude lower than that observed in our experiment. Similar discrepancies
between experimentally observed values of B$_{fj1}$ and those calculated within
the framework of the adiabatic theory, have been reported for other HTS system
\cite{nab97,wip92,mch92,ger93,gui89,gerb93,mll94}. In these studies, the
experimentally observed values of B$_{fj1}$ at 4.2 K are of an order of 1 T
or higher.

In order to account for why our sample appears more stable against flux jumping than
would be expected from adiabatic theory, two phenomena need be considered.  First, it
is possible that adiabatic conditions may simply not be fulfilled in our experiment.
Were this to be the case, we would expect the thermal anchoring of the sample to the cold
finger of the cryostat to strongly influence flux jumping. The second possibility is
flux creep. As was shown by McHenry and coworkers \cite{mch92}, flux creep stabilizes the
critical state of the superconductor against the flux jumps. Both the possibilities
may imply a dependence of the flux jumping on the sweep rate, as in fact is observed
experimentally. This means that decreasing the sweep rate may stabilize the
superconducting sample against flux jumping; thus the observed values of B$_{fj1}$
would be higher than those predicted by adiabatic theory.

Figure 6 shows the dependence of the field of the first flux jump,
B$_{fj1}$ on the magnetic field sweep rate. As the sweep rate
decreases the value of B$_{fj1}$ increases rapidly and at a sweep
rate of 0.12 T/min it approaches a value of about 4.5 T (Fig.6).
From Figure 6, one sees that B$_{fj1}$ tends to saturate at higher
sweep rates. The exact value of the B$_{fj1}$ field at the
saturation cannot be determined, as the maximum sweep rate
attainable in our system is 1.2 T/min. However, we estimate it to
be around 1 T, several times higher than the value predicted by
adiabatic theory. Similar strong sweep rate dependence on
B$_{fj1}$ was observed in other HTS systems
\cite{mch92,ger93,gerb93,gui88}.

To check whether the adiabatic conditions are satisfied in our experiment we need to
determine the relation between thermal (D$_t$) and magnetic (D$_m$) diffusivity.
These parameters are estimated by applying standard procedure presented in
reference \cite{mch92} from the formulas: D$_t$$\approx$$\frac{\kappa}{c}$ and
D$_m$$\approx$$\frac{\rho}{\mu_0}$,
where: $\kappa$ - thermal conductivity, c - specific heat and $\rho$ - resistivity.
The estimation of these parameters in \BISCO is difficult for several reasons.
Firstly, the transport properties of this system are extremely anisotropic.  The value
of the inplane thermal conductivity for \BISCO is estimated to
be $\kappa$(4.2 K)$\approx$ 1 W/Km \cite{dam89}, which gives
 D$_t$(4.2 K)$\approx$9x10$^{-4}$ m$^2$/s.
However, the value of the  thermal conductivity along the c-axis of \BISCO is
difficult to determine
experimentally, although one expects this value, as well as that for the thermal
diffusivity, to be significantly lower than in ab-plane, due to the planar nature of the
material.

The estimate of the magnetic diffusivity in superconductors is also difficult.
In nonsuperconducting materials the coefficient of the magnetic diffusivity is
proportional to their resistivity. The normal state resistivity in \BISCO is extremely
anisotropic \cite{mar88}. The in plane resistivity $\rho_{ab}$ varies linearly with
temperature with d$\rho_{ab}$/dT=0.46 $\mu$$\Omega$cm/K, whereas resistivity along the
c-direction $\rho_c$ is about five orders of magnitude higher \cite{mar88}. In the
configuration appropriate to our experiment, flux front moves in the direction parallel
to ab plane.
Hence, for the magnetic diffusivity, the estimation of $\rho_{ab}$ is the appropriate one.
Assuming a linear temperture dependence to $\rho_{ab}$ below T$_c$, one
obtains $\rho_{ab}$(4.2 K)$\approx$1.9x10$^{-8}$ $\Omega$m and therefore
D$_m$$^{ab-normal}$(4.2 K)$\approx$1.5x10$^{-2}$m$^2$/s for the normal state
diffusivity. This value of magnetic diffusivity is more than one order of magnitude
higher than the in plane thermal diffusivity. However, since our sample is not in a
normal state but in a superconducting mixed state, the flux flow resistivity (
$\rho_f$=$\rho_n$$\frac{B}{B_{c2}}$, where $\rho_n$ is the normal state resistivity,
and $B_{c2}$ is the upper critical field of the superconductor) must be used to
estimate the magnetic diffusivity.
On the basis of our magnetization data and above expressions, we estimate
$\rho_f$$\approx$$\frac{1}{30}$$\rho_n$ just before appearance of the first flux jump,
which gives D$_m$$^{ab}$(4.2 K)$\approx$5x10$^{-4}$ m$^2$/s. This value of magnetic
diffusivity is the same order of magnitude as the in plane thermal diffusivity. Therefore
adiabatic conditions may be not fulfilled in our experiment at 4.2 K and this may be
one reason for enhanced stability of our sample against flux jumping at this
temperature. We may also expect an influence of the heat exchange conditions on flux
jumping in our system as well. We should bear in mind, however, that our sample's
geometry is that of a relatively thin pellet with its flat surface parallel to the
ab-planes. Such a geometry minimizes heat exchange by transport within the ab-plane,
and heat transport along the c-axis is expected to be significantly lower than that in the
ab-plane.  On the other hand, the estimation of the magnetic diffusivity  on the
basis of flux flow resistivity alone, may be  questionable.
Recently, it was suggested by Mints \cite{dmints96} that the presence of  flux creep may
cause D$_m$$<<$D$_t$ and as a result the local heat exchange conditions to be
strongly non-adiabatic.

Flux creep is a phenomenon which may stabilize our sample
against flux jumping, as well as imply a sweep rate dependence on flux jumps.
\BISCO crystals are characterized by relatively strong magnetic relaxation
caused by  flux creep \cite {biggs89,sved91}. At low temperatures and for a relatively
small time window, this magnetic relaxation is logarithmic in time.
Similar magnetic relaxation was found in our sample (see Fig.3).
On the basis of our results we have estimated the effective pinning potential
(U$_{eff}$) to be U$_{eff}$/k$\approx$210 K at T=4.2 K and H=2 T, where k is the
Boltzman constant. This value of U$_{eff}$ is typical for the given temperature and
magnetic field range in \BISCO system.

Another system characterized by relatively strong flux creep and magnetic relaxation
is La$_{1-x}$Sr$_{x}$CuO$_{4}$ \cite{mchenry91}. Similarly, a strong influence of the
external magnetic field sweep rate on flux jumping was found in
La$_{1.86}$Sr$_{0.14}$CuO$_4$ crystals \cite{mch92}.
The presence of  flux creep
changes the magnetic flux profile in a superconducting sample during a sweep of an
external magnetic field. It was shown by McHenry and co-workers \cite{mch92} that in
the presence of flux creep, heat generated in the superconducting sample by a
fluctuation of an external magnetic field is relatively small compared to the case
when flux creep is absent. This fact influences the stability conditions of the
critical state. As a result we expect an increase of the value of B$_{fj1}$
relative to that predicted by the theory, in which  flux creep phenomenon is not
taken into account. Flux creep phenomenon can fully stabilize the superconducting
sample against flux jumping as the rate of magnetic field changes decreases.
Related behavior was indeed observed in our experiment (see Fig.2).

The influence of  flux creep on flux jumping was also analyzed theoretically by Mints
\cite{dmints96}. In this model a logarithmic dependence of the screening current
density on the electric field (induced by external magnetic field changes) was
assumed, whereas the thermal conditions were assumed to be extremely non-adiabatic.
(i.e. $\tau$$_t$$<<$$\tau$$_m$ or D$_t$$>>$D$_m$). As a result the predicted
values of the B$_{fj1}$ depend strongly on the heat transfer coefficient.
Assuming the Bean model \cite{bean62} (i.e. J$_c$(B)=const) this theory
predicts B$_{fj1}$$\sim$(dH$_e$/dt)$^{-1/2}$, where dH$_e$/dt is the
external magnetic sweep rate. A fit of this formula to our experimental data is
shown in Figure 6 by the dotted line. Comparison of the experimental and
the fitt curve show that while the experimental curve tends to
saturate at higher sweep rates around the value of B$_{fj1}$$\approx$1 T the formula
predicts that it approaches zero. It was shown in reference \cite{dmints96} that taking
into account a decrease of the critical current density with magnetic field slows down
the decrease of B$_{fj1}$ with increasing  dH$_e$/dt. Nevertheless, the
theoretical curve still approaches zero at sufficiently high sweep rates. For more
accurate comparison of the theoretical formula (from reference \cite{dmints96}) with
experiment, investigations at higher sweep rates would be necessary.
A quantitative comparison of this formula with our experimental results was
impossible, because neither the electric-field dependence of the screening current
nor the heat transfer coefficient in our experimental setup were known.

Finally, let us discuss the influence of demagnetizing effects on flux jumping. As
our sample is in a form of a thin pellet and the external magnetic field is aligned
perpendicular to its surface, one may expect a relatively strong influence of such
effects on the flux instabilities. The easiest way to analyze the influence of
demagnetizing effects on a magnetic material is to introduce a demagnetizing factor D.
In the case of superconductors this factor is often estimated from the initial slope
of the magnetization curve. Such procedure gives D$\approx$0.82 in our case. However,
the application of the demagnetizing factor is not fully appropriate in the case of
superconductors, due to the fact that the magnetization of superconductor is caused by
macroscopic screening currents. The self-component of the magnetic field significantly
alters the distribution of these currents in a superconducting sample. This fact was
confirmed by a number of experiments performed on both conventional \cite{frank79} and
high temperature superconductors \cite{theus92,tame93}. The experiments were performed
by magneto-optic technique \cite{theus92} or by using scanning
Hall-probes \cite{frank79, tame93} on thin conventional superconducting
discs \cite{frank79}, HTS films \cite{theus92} or thin single crystals
\cite{tame93}, in the external magnetic field perpendicular to their surfaces.
Until now, many attempts have been undertaken to solve the problem of the
magnetic field and screening current distribution in superconductors with non-zero
demagnetizing factors \cite{sanch01}. In most cases this problem cannot be solved
analytically and numerical calculations are necessary. However, the problem of
the distribution of the screening currents and magnetic field was solved analytically
for superconducting samples with large (close to 1) demagnetizing factors, i.e. for
samples in shape of infinitely long and thin strips \cite{bra93} or infinitely thin
disks \cite{dae89}.

Since the demagnetizing factor of our sample is large, let us discuss some features of
the model presented in reference \cite{bra93} pertinent to the present studies of flux
jumping.
In the above mentioned model, the magnetization of an infinitely long and thin strip
in an external magnetic field perpendicular to its surface as a function of this field
(after cooling the sample in zero field) is given by:

\begin{eqnarray}
\label{eq:three}
M&=&-{\frac{1}{2}}J_ca\tanh(\pi{\frac{\mu_0H_e}{B^{**}}})
\end{eqnarray}

where: $J_c$ is the critical current density, \textit{a} is half of the width of
the strip, $H_e$ is the external magnetic field,
B$^{**}$=$\mu_0$J$_c$d$^*$ and d$^*$ is the thickness of the strip.
This model assumes that: d$^*$$\ll$$\textit{a}$ and J$_c$(B)=const.

On the basis of equation (2) one can conclude that the magnetization of the
infinitely long and thin strip never saturates. The model implies that the center of
the strip is fully screened even for extremely large external magnetic fields. Hence,
one cannot use the concept of the field of full penetration the same way one would for
the well-known case of a sample with zero demagnetizing factor (for example an
infinitely long slab or cylinder). However, to some extent the role of the field of
full penetration in the case of a strip is assumed by the B$^{**}$ parameter. It is
important to note that in this case d$^*$ is the thickness of the sample i.e. the
dimension of the sample measured in the direction parallel to the external magnetic
field. In the case of our sample, d$\approx$0.2 mm and it is significantly smaller than
other sample dimensions. Using the critical current density we estimated earlier, we may
estimate B$^{**}$(2K)$\approx$5 T and B$^{**}$(6K)$\approx$3.5 T. These values are indeed
quite close to the maximum value of B$_{fj1}$ found in our experiment
i.e. 4.86 T at 6 K (Fig. 1). Thus, one may conclude that in the case of very thin
samples, such as single crystals or textured samples of the \BISCO system, with an
external magnetic field perpendicular to their surfaces, the role of the critical
dimension from the point of view of appearance of flux jumps, is played by the
thickness of the sample.

%\section{SUMMARY}
\section[SUMMARY]{SUMMARY}
Flux jumps phenomena in a thin \BISCO  textured sample with an external magnetic field
parallel to the c-axis have been systematically studied. The values of the
experimentally observed instability fields depend strongly on the external magnetic
field sweep rates. Both flux creep and the heat exchange conditions between the sample
and its environment must be taken into account in a quantitative analysis of this
phenomenon. For thin samples in an external magnetic field perpendicular to their
surfaces, the role of the critical dimension from the point of view of appearance of
flux jumps is played by their thickness.

We conclude that the necessary requirements for the avoiding flux jumping in \BISCO
single crystals are following:

(i) The sample thickness has to be lower than its critical dimension, of the order of
0.1 mm, when J$_c$ at the relevant field strengths and temperatures is of the order
of 10$^6$ A/cm$^2$.

(ii) The magnetic sweep rates have to be low enough to allow for the stabilizing
influence of flux creep to operate effectively and to allow for the effective heat
exchange between the sample and the experimental environment.

Most experiments performed so far on \BISCO single crystals reported in the literature
have been carried out on SQUID systems with very low magnetic sweep rates and on very
thin (of the order of 0.1 mm) samples. It is thus not surprising that flux jumps have
not been reported in these experiments.

\begin{acknowledgments}

We acknowledge financial support from the Natural Sciences and Engineering Research
Council of Canada and Polish Government Agency KBN contract No. 8T11B03817.
One of us (AN) would like to acknowledge a NATO Science Fellowship.
\end{acknowledgments}

\end{document}